\documentclass[11pt]{article}
\usepackage{amsmath}
\usepackage{amssymb}
\renewcommand{\title}[1]{%
    \bigskip%
    \begin{center}%
    \Large\bf #1%
    \end{center}%
    \vskip .2in}

\renewcommand{\author}[1]{%
    {\begin{center}
    #1
    \end{center}}}
\newcommand{\address}[1]{\vspace{-1.7em}\vspace{0pt}
    {\begin{center}
    \it #1
    \end{center}}}

\begin{document}

\title{\bf{Canonical Formulation of a New Action for a Non-relativistic Particle Coupled to Gravity}}

\author
{
Rabin Banerjee  $\,^{\rm a,b}$,
Pradip Mukherjee  $\,^{\rm c,d}$}
\address{$^{\rm a}$ S. N. Bose National Centre 
for Basic Sciences, JD Block, Sector III, Salt Lake City, Kolkata -700 098, India }

\address{$^{\rm c}$Department of Physics, Barasat Government College,Barasat, India}
\address{$^{\rm b}$\tt rabin@bose.res.in}
\address{$^{\rm d}$\tt mukhpradip@gmail.com}

\vskip 1cm
\begin{abstract}
A detailed canonical treatment of a new action for a nonrelativistic particle coupled to background gravity, recently given by us \cite{BM8}, is performed both in the Lagrangian and Hamiltonian formulations. The equation of motion is shown to satisfy the geodesic equation in the Newton-Cartan background, thereby clearing certain confusions. The Hamiltonian analysis is done in the gauge independent as well as gauge fixed approaches, following Dirac's analysis of constraints. The physical (canonical) variables are identified and the path to canonical quantisation is outlined by explicitly deriving the Schroedinger equation. Usual flat space results are correctly reproduced.

\end{abstract}


\section{Introduction }

\smallskip

 Newton Cartan (NC) space time is a four dimensional differentiable manifold with two degenerate metrics. Just after Einstein formulated the general theory of relativity (GR) as a relativistic theory of gravity, Cartan demonstrated that Newtonian gravity can also be formulated as a geometric theory in  NC manifold  \cite{Cartan-1923, Cartan-1924}. Intense research on  various aspects of the metric theory produced a rich literature 
 \cite{Havas, Daut, TrautA, Kuch, EHL}. The metric properties of the NC space time being so different from the Riemann or Riemann Cartan space time that great care had to be taken 
to derive  Newtonian gravity from General relativity. In case of the Riemann space there is a unique non singular metric but in NC geometry there are two { degenerate} metrics. A direct outcome is the difficulties  in  coupling of matter theories with non relativistic gravity. In the 'classical' age this issue was not very prominent.

  Resurgence of this field in recent times, is due to  the applications of the geometric approach to physical
 phenomena in varied topics including  condensed matter physics, hydrodynamics and cosmology. 
 So the question of coupling matter systems with non relativistic gravity occupied the centre stage.
 A number of different approaches to the problem have appeared  in the recent past \cite{SW, j1, ABPR}, the most popular among these is based on the gauging of 
 (extended) Galilean group algebra \cite{ABPR, PPP}. In this scenario we bave developed 
a systematic algorithm, developed by us in a series of papers \cite{BMM1, BMM2, BM4},  of gauging the relevant Galilean symmetries. It was  generically termed Galilean gauge theory {\bf  (GGT)} in analogy with Poincare gauge theory that is obtained by gauging the Poincare symmetries. 
  In this paper we develop a canonical formalism for a nonrelativistic particle coupled to background  { NC} geometry. The action for this theory was very recently derived by us \cite{BM8} usiing GGT.

  The advantage of our theory is
the obtention of the geodesic equation which is nontrivial since earlier approaches were not successful in deriving the geodesic equation \cite{PPP}. The last method is an example o gauging the algebra There is another approach where the generator of the central charge of the projective Galilean algebra appears in addition to the generators of the algebra that are dynamical. Only the later fields
are gauged in GGT. This difference leads to different notions of flat limit. In the case of GGT the flat limit  is a free particle whereas in the second case the flat limit corresponds to an other wise free particle moving under a Newtonian potential. We will point out these issues at the appropriate place in the following.

 The Lagrangian description is backed by a Hamiltonian analysis. There is a single  first class constraint that is shown to generate the reparametrisation symmetry. The gauge independent analysis is supplemented by a gauge fixed formulation where the raparametrisation parameter is identified with the absolute time. We show that it is a good choice of gauge since it allows the abstraction of the physical (canonical) variables in a simple manner. The Dirac brackets of the canonical set are identical to the Poisson brackets which therefore allows us to elevate them to commutators, when quantising the theory. The operator version of the relevant variables are given in the coordinate representation. This allows us to write the Schroedinger equation, thereby paving the way for the quantisation of the model. In all cases the appropriate flat limit is reproduced.

It will now be appropriate to describe the organisation of the paper. The action mooted by us in \cite{BM8} is reviewed in the next section with emphasis on its connection with the Newton Cartan geometry. This is crucial to understand its difference with the earlier results \cite{PPP}. This discussion will be found in the the last subsection of section 2. One of the most important result of  section 3 is to show that the path of the particle is the standard geodesic if the parameter labelling  the path is an affine parameter. This result is obtained from the Lagrangian analysis given in section 3. This is followed by the Hamiltonian analysis in section 4. Since the system is singular
Dirac's approach of constrained Hamiltonian analysis \cite{D} has been used. We show that there is
one first class constraint. Since the system is also generally covariant, the canonical Hamiltonian vanishes. We  also provide a gauge fixed analysis imposing the static gauge.  The Dirac brackets
paved the way towards quantization. Section 4 contains this comprehensive Hamiltonian analysis leading to the Schroedinger equation for the system. Finally, we conclude in section 5.

\smallskip

\section{Action for Non relativistic Particle in Curved Background}
 
  The  parametrized action for a non relativistic  particle in $3$ dimensional Euclidean space and absolute time
  is given by,
\begin{equation}
S = 
\int
\dfrac{1}{2}m\dfrac{ \dfrac{dx^{a}}{d\lambda}\dfrac{dx^{a}}{d\lambda}}{\bigg(\dfrac{dx^{0}}{d\lambda}\bigg)} ~d\lambda
\label{actionum}
\end{equation}
where the index $a$ denotes a space index. The action is invariant under the global Galilean transformations,
 \begin{equation}
x^\mu \to x^\mu + \xi^\mu ; \xi^0 =  -\epsilon, \xi^k = \eta^k - v^k t; \eta^k= \omega^k{}_l x^l+ \epsilon^k\label{globalgalilean} 
\end{equation}

  This invariance is ensured by the transformations

 \begin{equation}
\delta \dfrac{dx^{0}}{d\lambda} = \dfrac{d}{d\lambda}(\delta x^{0}) = - \dfrac{d\epsilon}{d\lambda} = 0
\label{der1}
\end{equation}
as $\epsilon $ is constant and,

\begin{equation}
\delta \dfrac{dx^{k}}{d\lambda} = w^{k}_{j}\dfrac{dx^{j}}{d\lambda} - v^{k}\dfrac{dx^{0}}{d\lambda}
\label{der2x}
\end{equation}
which can be checked easily. The Lagrangian (\ref{actionum}) changes by,
\begin{equation}
\delta L = -\dfrac{d}{d\lambda}\bigg(mv^{k}\dfrac{dx^{k}}{d\lambda}\bigg)
\label{variation}
\end{equation}
due to (\ref{der1}) and (\ref{der2x}).
The change of the action (\ref{actionum}) is then a boundary terms only,
see (\ref{variation}). The same equations of motion follow from both the original and the transformed action. So the theory is invariant under the global Galilean transformations.

 To localize the symmetry of the action (\ref{actionum}) according to GGT ,$\frac{dx^{\alpha}}{d\lambda}$ is now subsituted by the covariant derivatives $\frac{Dx^{\alpha}}{d\lambda}$, where
\begin{equation}
\dfrac{Dx^{\alpha}}{d\lambda} = \dfrac{dx^{\nu}}{d\lambda} \Lambda^{\beta}_{\nu}\partial_{\beta}x^{\alpha} = \dfrac{dx^{\nu}}{d\lambda} \Lambda^{\alpha}_{\nu}
\label{red}
\end{equation}
Here $\Lambda^{\beta}_{\nu}$ are a set of new compensating (gauge) fields. Note that 
 the localisation procedure can be tuned smoothly to restore global Galilean symmetry. In this limit the covariant derivatives $\frac{Dx^\alpha}{d\lambda}$ must then go to the ordinary derivatives $\frac{d x^\mu}{d\lambda}$.  This sets the following condition 
\begin {equation}
\Lambda^\alpha{}_\mu \longrightarrow \delta^\alpha _\mu
\label{NS}
\end {equation}
This at once shows that $ \Lambda^\alpha{}_\mu $ is non singular i.e. the corresponding matrix is invertible. We will further see that this observation is instrumental in the geometric interpretation of our theory.

 Now the transformations of the new gauge fields should  ensure
that the 'covariant derivatives' will transform under the local Galilean transfrmations in the same form as the usual derivatives do under the global Galilean transformations. Then the new theory obtained by replacing the ordinary derivatives by the `covariant derivatives' will be invariant under the local Galilean  transformations. This is the essence of GGT \cite{BMM1, BMM2}. Using (\ref{der1}) and (\ref{der2x}) the  transformation of the covariant derivatives follows,

 \begin{equation}
\delta \dfrac{Dx^{0}}{d\lambda} = 0
\label{covder12}
\end{equation}
and, likewise for the space part,

\begin{equation}
\delta \dfrac{Dx^{k}}{d\lambda} = w^{k}_{j}\dfrac{Dx^{j}}{d\lambda} - v^{k}\dfrac{Dx^{0}}{d\lambda}
\label{covder2n}
\end{equation}

Exploiting these relations the transformations of the newly introduced fields are completely specified. They are given by \cite{BM8},

\begin{eqnarray}
\delta \Lambda_0^0 &=& \dot\epsilon \Lambda_0^0 \nonumber\\
\delta \Lambda_i^a &=& \omega_b^a \Lambda_i^b - \partial_i\xi^k \Lambda_k^a \nonumber\\
\delta \Lambda_0^a &=& \dot\epsilon \Lambda_0^a - v^a\Lambda_0^0 - \partial_0 \xi^k \Lambda_k^a + \omega_b^a \Lambda_0^b
\label{old1x}
\end{eqnarray} 
while the remaining field $\Lambda_i^0$ simply vanishes.

 Hence the cherished action is given by,
\begin{equation}
S = 
\int
\dfrac{1}{2}m\dfrac{ \dfrac{Dx^{a}}{d\lambda}\dfrac{Dx^{a}}{d\lambda}}{\bigg(\dfrac{Dx^{0}}{d\lambda}\bigg)} ~d\lambda
\label{actionm}
\end{equation}
obtained from (\ref{actionum}), substituting $\frac{dx^\mu}{d\lambda}$ by $\frac{Dx^{\mu}}{d\lambda}$.  
  The above derivation is the standard procedure of GGT which leads to the theory (\ref{actionm}) {\it in flat space} invariant under transformations formally similar to (\ref{globalgalilean}). But the parameters of transformations. But they are not constants and vary in a special manner with space and time, completely in unison with the privilaged role of time.in Newtonian theory.
  
\subsection{Geometrical Connection}

  The GGT has a remarkable feature. The transformations of the gauge fields introduced during localisation 
have a suggestive form which is exhibited by  the set of transformations obtained for a generic model
\cite{BM4} having Galilean symmetry. The transformations obtained here (\ref{old1x}) are no exceptions. The form variation of the gauge fields carry two indices, a local and a global one. If we reinterpret the theory as a theory in curved space time, the local indices transform as under Galilean rotations and boosts while the global indices resemble a diffeomorphism
$x^\mu \to x^\mu + \xi^\mu$.

The new fields ${\Lambda_\mu}^{\alpha}$ may then be reinterpreted as inverse vielbeins in a general manifold charted by the  coordinates $x^{\mu}$ connecting the (global) spacetime coordinates with the local coordinates. A crucial requirement for this is the existence of an inverse which is met above by the continuity of the local to global invariances  (see the equation (\ref{NS}) and the discussion around it )\footnote{A word about our notation: indices from the beginning of the alphabet ($\alpha,  \beta$ etc. or $a, b $ etc) denote the local basis while those from the middle ($\mu, \nu$ etc. or $i, j$ etc.) indicate the global basis; Greek indices denote space-time while only space is given by the Latin ones. Repeated indices denote a summation.} The geometric reinterpretation is then seen from the above transformations (see, for example (\ref{old1x}))  where the local indices (denoted by $a$) are Lorentz rotated, while the global indices (denoted by $i$) are coordinate transformed. 

Before proceeding to the next discussion note that there is no convective term in the transformation of
 $\Lambda_i^a$. This is because the variations involve only changes in the form,
\begin{equation}
\delta\Lambda_i^a =\Lambda_i^{\prime a} (\lambda) -\Lambda_i^a(\lambda)
\label{formvariation}
\end{equation}
where $\lambda$ is the parameter that does not change. This happens when we do particle mechanics. A similar structure appeared earlier in (\ref{NS}), for the variation of the spatial metric $h_{ij}$,  that was discussed in \cite{PPP}. For field theory, the convective term appears as both the form of the variable as well as the coordinates are changed. As an example, 
\begin{equation}
\delta g_{ij} = g_{ij}'(x') - g_{ij} (x)
\label{totalvariation}
\end{equation}
To distinguish from the form variation in (\ref{formvariation}), the above is called the total variation in the literature.

 Similar conclusions hold for the other two transformations. 
It has been proved that the 4-dim space time obtained in this way above is the Newton-Cartan
manifold. This is done by showing that the metric formulation of our theory contains the same structures and satisfy the same structural relations as in NC space - time \cite{BMM2}. 
 As is well known that the NC manifold is a degenerate one. The metric properties is fully expressed by a singular matrix
$h^{\mu\nu}$ and a one form $\tau_\mu$ which defines the absolute direction of the flow of time \footnote{Thus this one form determines the foliation of space time, where the Newtonian dynamical variables are defined on the slices The coordinate system defined using this foliation are the Galilean (adapted) coordinates. The most important advantage of the GGT is that it automatically 
couples the free theory in the Galilean coordinates.}. The Newton Cartan geometry is a degenerate manifold with a singular metric $h^{\mu\nu}$ and a one form $\tau_\mu $ satisfying the following 
algebra,
\begin{eqnarray}
h^{\mu\nu} \tau_\nu = 0 \quad\quad  \tau ^\mu \tau_\mu = 1\nonumber\\
h_{\nu\mu} h^{\mu\rho} = P^\rho _\nu =\delta^\rho _\nu - \tau^\rho\tau_\nu
\label{algebra}
\end{eqnarray}
where, $P^\rho _\nu$ is the projection  operator. The Milne symmetry of NC geometry is once again a cardinal issue in the field. That this symmetry is connected with the local Galilean boost was  known earlier. But a comprehensive theory of this connection was obtained from GGT \cite{BM5}.  The quantities  $h_{\mu\nu}$ and $\tau^\mu$ are additional structures defined for raising or lowering indices.

 With this short introduction to the Newton Cartan algebra we ask the question  whether our candidates for the vielbeins and their inverse may be used to build up the metric structures with appropriate
 transformations and satisfy the set (\ref{algebra})?
 
  We begin with the metrics. One can
define  
   
 \begin{equation}
h^{\mu\nu}={\Sigma_a}^{\mu}{\Sigma_a}^{\nu}; \hspace{.2cm}\tau_{\mu}={\Lambda_\mu}^{0} =\Theta\delta_\mu^0
\label{spm}
\end{equation}
and,
\begin{equation}
h_{\nu\rho}=\Lambda_{\nu}{}^{a} \Lambda_{\rho}{}^{a}; \hspace{.2cm}\tau^{\mu}=
{\Sigma_0}^{\mu}\hspace{.3cm}
\label{spm2}
\end{equation}
where $\Sigma_\alpha{}^\nu $ is the inverse of ${\Lambda_\mu}^{\alpha}$  , 
\begin{equation}
\Sigma_\alpha{}^\nu{\Lambda_\nu}^{\beta} =\delta^\beta_\alpha \hspace{.2cm};\hspace{.2cm}
\Sigma_\alpha{}^\nu{\Lambda_\mu}^{\alpha} =\delta^\nu_\mu \label{B}
\end{equation}

Knowing the transformations of the $\Lambda's$ it is possible to compute the transformations of the $\Sigma's$ from the above relations \cite{BMM2}. The local and global basis are thus appropriately connected by these vielbeins in the following way,

\begin{equation}
\hat e_\mu = \Lambda_\mu^\alpha\hat e_\alpha, \,\,\, \hat e_\alpha = \Sigma_\alpha^\mu \hat e_\mu
\label{basis}
\end{equation}

For flat spacetime, there should be no difference between the local and global bases. This has been already established in (\ref{NS}) as the vielbeins simplify to Kronecker deltas in the flat limit. Since the vielbeins are connected to the identity, these are obviously invertible. Moreover, we have already shown that the appropriate tensorial properties are satisfied by them. Hence if the findings hold for any specific coordinate system, they will also hold for any other system which shows the gauge independence of the results. Thus we are free to choose the particular coordinate system where we perform our analysis.

All the above relations are  trivially satisfied in the flat space limit.  A slightly non trivial check is provided by the flat space limit of (\ref{old1x}). The transformations parameters are now spacetime independent (ordinary global Galilean parameters). While the left side vanishes trivially, the right vanishes on using  (\ref{globalgalilean}).

We may remind that $h^{\mu\nu}$ is not the inverse of $h_{\mu\nu}$, being the elements of Newton Cartan geometry, satisfying (\ref{NS}). However, since rotational symmetry is preserved $h^{ij}$ is the inverse of $h_{ij}$, i.e.,
\begin{equation}
h^{ik} h_{kj} = \delta ^i_j
\end{equation}
which may also be verified from (\ref{NS}). 

 Thus $h_{ij}$ may be regarded as a  non degenerate spatial metric which should transform as the space-space components of a rank 2 covariant tensor.
 This is proved from our relations. Note that
the parameters $\xi^\mu$ are now local (i.e. space time dependent), except for the time translation parameter $\epsilon$ which, as stated earlier, is a function of time only, keeping in view the absolute nature of time in nonrelativistic physics. Under infinitesimal transformations we should have,

\begin{equation}
\delta h_{\mu\nu} = - h_{\alpha\nu}\partial_\mu\xi^\alpha -
 h_{\alpha\mu}\partial_\nu\xi^\alpha
\label{htransform}
\end{equation}
 Taking the space space component we find,

\begin{equation}
\delta h_{ij} = - h_{kj}\partial_i\xi^k - h_{ki}\partial_j\xi^k
\label{proof}
\end{equation}
where the $\alpha=0$ components drops out since $\xi^0$, the time translation parameter, is a function of time only. Note that as usual there is no convective term present in the above transformation law, the reason for which was discussed at the beginning of section 3.1. In fact it is exactly identical to that given in (\ref{NS}), that was discussed in \cite{PPP}. From (\ref{spm2}) and (\ref{old1x}), we find,
\begin{equation}
\delta h_{ij} = (\omega_b^a \Lambda_i^b - \partial_i \xi^k \Lambda_k^a)\Lambda_j^a + (\omega_b^a \Lambda_j^b - \partial_j \xi^k \Lambda_k^a)
\Lambda_i^a 
\label{proof1}
\end{equation}
Terms involving the rotation parameter cancel from symmetry arguments. The remaining pair of terms are simplified by using (\ref{spm2}). It immediately reproduces (\ref{proof}).

It is now easy to express (\ref{actionm}) in a manifestly covariant form using the Newton Cartan elements. From (\ref{red}) and (\ref{spm2}) we obtain,
\begin{equation}
\dfrac{Dx^{a}}{d\lambda}\dfrac{Dx^{a}}{d\lambda} = h_{\nu\sigma}\dfrac{dx^{\nu}}{d\lambda} \dfrac{dx^{\sigma}}{d\lambda} 
\label{red1}
\end{equation}
so that the action may be written as,
\begin{equation}
 S=  \left(\dfrac{m}{2}\right)\int h_{\nu\rho}\frac{{x'^\nu} x'^{\rho}}{\Theta x'^0} ~d\lambda
\label{lagm10}
\end{equation}
 Clearly, this  can be interpreted
as the action of a non relativistic  particle coupled with a Newton Cartan background.\footnote{We work with zero torsion.}.

\subsection{Comparison with earlier results } 
Here we undertake a comparison of our action with the earlier results. Meanwhile let us note that
The action (\ref{lagm10}) is quite satisfactory.
 The flat limit poses no problems. In this limit we recall that the vielbeins reduce to the Kronecker deltas. Then $\Theta =1$, $h_{00}=h_{0i}=0$ and $h_{ij}=\delta_{ij}$. The action (\ref{lagm10}) reduces to the standard NR action for a free particle in flat space, expressed in a reparametrisation invariant form. 
 
 In other approaches \cite{PPP} the form of the action is not the same as (\ref{lagm10}). Apart from a term which is similar to (\ref{lagm10}) there occurs another piece that depends on a gauge field. Thus the two representations appear to be inequivalent But, this is not so. The apparent paradox is resolved by looking at the flat space limit. As shown above, the action (\ref{lagm10}) passes to the NR flat space action (\ref{lagm1}). On the contrary the action obtained from the other approaches reduces to (\ref{lagm1}) plus a term which is interpreated as a potential which is a solution of a Poisson equation. It comes from the gauge field part of the usual action. This connection was explained in \cite{MPR}. 
 
 It is now clear why we get an action that is different from the action obtain from other approaches. Our results cruucially depends on the defiinition 
 of the flat space theory. The (global) Galiilean symmetry of the flat theory are gauged (localised) to eventually yields a curved generalisation of the flat space theory. The flat space theory may be free, as is considered here. Or, it may be interacting as we have considered elsewhere \cite{BMM3, SW}interdiction, it is crucial to note, must be non gravitational. Thus it will not be possible to start from a theory that contains a gravitational potential.
 
 One may wonder whether it is still might possible to smuggle the gauge field in our expression (\ref{lagm10}) by using the transformations under the Milne boost symmetry .This can not be done Consider the Milne boost transformation
 
 
  
  \begin{eqnarray}
  h_{\mu\nu} =  {\bar{h}}_{\mu\nu} + \left( {\bar{\tau}}_\mu P_\nu^\rho +{\bar{\tau}}_\nu P_\mu^\rho  \right) A_\rho - \left({\bar{\tau}}_\mu {\bar{\tau}}_\nu {\bar{h}}^{\rho\sigma}A_\rho A_\sigma \right)
  \nonumber\\
  \tau^\nu = {\bar{\tau}}^\nu \quad   \tau_\nu = {\bar{\tau}}_\nu \quad\quad {\bar{h}}^{\mu \nu} =h^{\mu\nu}
  \label{Milne}
   \end{eqnarray}
  which will connect the two. The quantity $P_\nu ^\rho$ is the projection operator, already defined in (\ref{algebra}).
  
   and $A_\mu$
is an arbitry one form.
Then the action changes to



  \begin{eqnarray}
\bar{ S} =  \left(\dfrac{m}{2}\right)\int ~d\lambda \left[ \bar {h}_{\nu\rho}\frac{{x'^\nu} x'^{\rho}}{\Theta x'^0}
+ 2\Theta x^{\prime 0} x^{\prime \nu}A_\nu - 
2\Theta^2 x^{\prime 0}{}^2 \tau ^\rho A_\rho +
\Theta ^2 x^{\prime 0} h^{\rho\alpha} A_\rho A_\alpha \right]
\label{lagm101}
 \end{eqnarray}  
  
  To extract the physical significance of (\ref{lagm101}) let us take its flat limit. In this situation
   \begin{eqnarray}
  \Theta = 1, \quad\quad h^{00} = h_{0 \mu}= 0\quad\quad h_{ij} = \delta{ij }\quad\quad A_{\mu}
  = \left(\phi, 0, 0, 0\right)
  \end{eqnarray} we get the free particle action (\ref{actionum}). The Newtonian gravitational potential $\phi$ does not appear. In fact this little exercise shows the consistency of our approach by reprodcing the orignal flat action, proving that the occurrence of $A_\mu$ in (
  \ref{lagm101}) is a gauge artifact of the Milne  symmetry. 
  
  It may be recalled that the gauge field may enter the structure of the NC geometry through the
arbitrary two form that occurs in the definition of connection, using the Trautman - Ehler 's conditions.
In our case the two form vanishes (see the discussion below (\ref{g1})) so that such a probability is ruled
out. This is a reassuring point compatible with our previous analysis.
 

\section{Lagrangian Analysis }

We start from the following action  of the non relativistic particle in curved background,
\begin{equation}
 S=  \left(\dfrac{m}{2}\right)\int h_{\nu\rho}\frac{{x'^\nu} x'^{\rho}}{\Theta x'^0} ~d\lambda
 = \left(\dfrac{m}{2}\right)\int h_{\nu\rho}\frac{{x'^\nu} x'^{\rho}}{\tau_\sigma  x'^\sigma} ~d\lambda
\label{lagm1}
\end{equation}
where, as earlier stated, a prime denotes a differentiation with respect to the parameter $\lambda$. The passage from the first to the second equality follows on using the  representation of $\tau_\sigma$. Note that the gauge fields do not appear explicitly in the above action, they have been absorbed in the definitions of the Newton Cartan structures $h_{\mu\nu}$ and $\tau_\mu$. Obviously it is different from the result needing an explicit introduction of the $U(1)$ gauge field. In the particular case obtained by setting $\phi = 0$ there, it has a structural similarity with our action. Our result is more in line with that of \cite{Kuch}. The total hamiltonian, obtained from a canonical analysis of the above action, which has been done later, agrees with the super-hamiltonian of \cite{Kuch}.

This action is manifestly invariant under the finite reparametrisations,
\begin{equation}
\lambda\to \lambda',  \,\,\, x^\mu(\lambda)\to x'^\mu(\lambda')
\label{reparametrisation}
\end{equation}
The  infinitesimal version is given by,
\begin{equation}
\lambda' =  \lambda + \delta\lambda, \,\,\, \delta x^\mu(\lambda) = \delta\lambda \frac{dx^\mu}{d\lambda}
\label{infinitesimal}
\end{equation}

 The Euler- Lagrange equation following from (\ref{lagm1})  is,
 \begin{eqnarray}
 \frac{d}{d\lambda}\left(\frac{\partial L}{\partial x'^\mu}\right) - \frac{\partial L}{\partial x^\mu}  =0 
 \end{eqnarray}
 
 We will give the calculations in some detail.  The derivatives of $L$ can 
 be straightforwardly computed. Multiplying the overall equation by $h^{\omega\mu}$ we get,
 
 \begin{eqnarray}
 x^{\prime \prime \omega} + \left( \tau^\prime . x^\prime \right)\tau^\omega - \frac{{ \left( \tau . x^\prime \right)}^\prime}
 { \left( \tau . x^\prime \right)} x^
 {\prime \omega} &-& \frac{h^{\omega\alpha}\tau^\prime{}_\alpha h_{\rho\beta} x^{\prime\rho}  x^{\prime  \beta} }{2\left( \tau . x^\prime \right)} + \frac{h^{\omega\alpha}
 h_{\mu\nu} x^{\prime\mu}  x^{\prime  \nu}\left(\partial_\alpha \tau_\sigma \right)x^{\prime  \sigma} }{2\left( \tau . x^\prime \right)}\nonumber\\
 & + & \frac{h^{\omega\alpha}}{2}\left(\partial_\sigma h_{\alpha \beta}
 +\partial_\beta h_{\alpha \sigma} -\partial_\alpha h_{\sigma \beta}\right) x^{\prime \sigma}  x^{\prime  \beta} = 0
 \label{intermediate}
 \end{eqnarray}
where the abbreviation,
\begin{equation}
\tau.x=\tau_\sigma x^\sigma
\end{equation}
has been used.
 
 We can now introduce the Dautcourt connection,
 \begin{eqnarray}
 \Gamma^\omega{}_{\sigma\beta} = \frac{1}{2} \tau^\omega \left(\partial_\sigma \tau_{ \beta}
+ \partial_\beta \tau_\sigma \right) + \frac{h^{\omega\alpha}}{2}\left(\partial_\sigma h_{\alpha \beta}
 +\partial_\beta h_{\alpha \sigma} -\partial_\alpha h_{\sigma \beta}\right)+ \frac{1}{2} h^{\omega\alpha} \left(K_{\alpha\sigma} \tau_\beta
+ K_{\alpha\beta} \tau_{ \sigma} \right)
\label{d} 
 \end{eqnarray}

 where $K$ is an arbitrary two form.
Now from (\ref{d}) we can write
\begin{eqnarray}
  \frac{h^{\omega\alpha}}{2}\left(\partial_\sigma h_{\alpha \beta}+
 \partial_\beta h_{\alpha \sigma} -\partial_\alpha h_{\sigma \beta}\right) x^{\prime \sigma}  x^{\prime  \beta} = \Gamma^\omega{}_{\sigma\beta} x^{\prime \sigma}  x^{\prime  \beta}-\tau^\omega \partial_\sigma \tau_{ \beta}  x^{\prime \sigma}  x^{\prime  \beta}-  h^{\omega\alpha} K_{\alpha\sigma} \tau_\beta  x^{\prime \sigma}  x^{\prime  \beta}
\end{eqnarray} 
Using this and the identity
\begin{eqnarray}
\tau_\alpha^\prime - \partial_\alpha\tau_\sigma x'^\sigma =
\left(\partial_\sigma\tau_\alpha - \partial_\alpha\tau_\sigma\right)x'^\sigma
\end{eqnarray}
in (\ref{intermediate}) we get the path of a particle falling freely in background gravity,
 \begin{eqnarray}
 x^{\prime \prime \omega} +  \Gamma^\omega{}_{\sigma\beta} x^{\prime \sigma}  x^{\prime  \beta} = \frac{{ \left( \tau . x^\prime \right)}^\prime}{ \left( \tau . x^\prime \right)}x'^\omega
 \label{g1}
 \end{eqnarray}
 where we have identified the arbitrary two form as,
 \begin{eqnarray}
K_{\sigma\alpha} = \frac{1}{2 (\tau.x')^2} \left(\partial_\sigma\tau_\alpha - \partial_\alpha\tau_\sigma \right)h_{\rho\beta}x^{\prime\rho}x^{\prime\beta}
 \end{eqnarray}
  Since we are doing torsionless NC geometry the above two form vanishes
as $ (\partial_\sigma \tau_\alpha - \partial_\alpha\tau_\sigma) $ is just
the temporal components of the torsion tensor \cite{BM5}. Also, the path is the equation of a geodesic in Newton Cartan geometry. Thus our action (\ref{actionm}) produces the correct geodesic equation. Further, the  arbitrariness of Dautcourt formula for the symmetric connection is eliminated. { \it This does not mean that the arbitrariness in the NC affine connection is removed. It shows that such arbitrariness
does not affect the NR spinless particle dynamics}.

In order to derive the affine form of the geodesic where the right side of (\ref{g1}) vanishes, the affine parameter has to be identified. This is easily done. The affine properties of the Newton Cartan geometry is determined by the direction of flow of time. The Galilean frame assumed in this work has the time axis oriented along the direction of absolute time. Substituting $\lambda = t$ in (\ref{g1}) we get  
\begin{equation}
\frac{d^2 x^\mu}{dt^2} + \Gamma^\mu{}_{\sigma\beta}
\frac{dx^{\sigma}}{dt}
\frac{d x^\beta}{d t} = \frac{\dot{\Theta}}{\Theta} \dot{x}{}^\mu
\end{equation} 
where an overdot denotes time differentiation. 
Now to fix the scale of time define  the affine parameter $T$ by
\begin{equation}
dT = \Theta dt
\end{equation}
This leads to the affine geodesic equation
\begin{eqnarray}
\frac{d^2 x^\mu}{d T^2} + \Gamma^\mu{}_{\sigma\beta}
\frac{dx^{\sigma}}{dT}
\frac{dx^{\beta}}{dT}=0\label{geo}
\end{eqnarray}

We have successfully constructed the action for a NR particle coupled to Newtonian gravity following the systematic procedure provided by galilean gauge theory (GGT) \cite{BMM1, BMM2, BM4}.
The background space time is identified with the Newton Cartan space time. A Lagrangian analysis has shown that a freely falling particle follows a geodesic in this NC spacetime. 
 \section{Hamiltonian Formulation}
 
 We follow Dirac's method of constrained systems \cite{D}  to develop the Hamiltonian formulation. This can be done either in a gauge independent manner or by fixing a specific gauge  \cite {HRT}. We do in both ways.
 
 \smallskip
 
 \subsection {\it{Gauge independent analysis}}

The dynamical fields in the action are $x^\mu$. The vielbeins $\Lambda(x)$ are prescribed functions of $x$.
The momentum, canonically conjugate to $x^\mu$ are ,
\begin{equation}
p_\mu =\frac{\partial L}{\partial x'^\mu}=\frac{mh_{\mu\nu}x^{\prime \nu}}{\left( \tau. x^{\prime}\right)} - \frac{mh_{\rho\nu}x^{\prime \rho}x^{\prime \nu}\tau_\mu}{2\left( \tau. x^{\prime}{}^2\right)}\label{canp}
\end{equation}
From here, after a couple of manipulations,  we get a primary constraint,
\begin{eqnarray}
\Omega_1 = \tau^\mu p_\mu +\frac{1}{2m}h^{\rho\sigma}p_\rho p_\sigma \approx  0\label{pc1}
\end{eqnarray}
In flat space the only nonvanishing components are given by $\tau^0=1, h^{ij}=\delta^{ij}$, so that the constraint simplifies to the well known energy-momentum condition,
\begin{equation}
E=\frac{\bf{p}^2}{2m}
\label{nrconstraint}
\end{equation}
where the energy is identified as, $E=-p_0$.

Expectedly, the canonical Hamiltonian vanishes,
\begin{eqnarray}
H_c = p_\mu x^{\prime\mu } - L = 0
\end{eqnarray}
 which is a manifestation of  the reparametrization invariance of (\ref{actionm}) \cite{HRT}. The total Hamiltonian is then given by just the constraint,
\begin{eqnarray}
H_T =\chi \left( \tau^\mu p_\mu +\frac{1}{2m}h^{\rho\sigma}p_\rho p_\sigma \right)\label{pc11}
\end{eqnarray}
where $\chi$ is a Lagrange multiplier. Since the Poisson algebra of constraints is strongly involutive, 
\begin{equation}
\{\Omega_1, \Omega_1\}=0
\end{equation}
there are no further constraints. We therefore have a single first class constraint which will subsequently be shown to generate the reparametrisation symmetry.

The Lagrange multiplier $\chi$  can be fixed from the canonical equation of motion,
\begin{eqnarray}
x^{\prime \mu} = \{ x^\mu, H\}_{PB}
\end{eqnarray}
Some calculation yields 
\begin{equation}
\chi = \tau. x^\prime
\end{equation}
 The total hamiltonian is then given by,
\begin{eqnarray}
H_T =\left( \tau. x^\prime\right) \left( \tau^\mu p_\mu +\frac{1}{2m}h^{\rho\sigma}p_\rho p_\sigma \right)\label{pc13}
\label{pc14}
\end{eqnarray}
This is exactly equal to the super Hamiltonian of Kucha\v{r} \cite{Kuch}.
  
 Since there is only one first class constraint, the gauge generator is just given by,
\begin{eqnarray}
G = \epsilon \Omega_1=\epsilon  \left( \tau^\mu p_\mu +\frac{1}{2m}h^{\rho\sigma}p_\rho p_\sigma  \right)   
\end{eqnarray} 
where $\epsilon$ is the gauge parameter.  

The change in $x^\mu$ is given by,
\begin{eqnarray}
\delta x^\mu = \epsilon \{x^\mu, \Omega_1\}=\frac{\epsilon}{\tau.x^\prime} x^{\prime \mu}\label{gx}
\end{eqnarray}
As already shown, 
the model (\ref{actionm}) has reparametrization invariance. 
 Comparing the above result  with (\ref{infinitesimal}) we find 
that the Hamiltonian gauge symmetry parameter is mapped to the  reparametrization parameter,
 \begin{equation}
 \delta\lambda = \frac{\epsilon}{\tau.x^\prime} 
 \label{comparison} 
 \end{equation}

We now calculate the Hamilton's equations of motion. These are given by bracketing with the total hamiltonian,

 \begin{eqnarray}
 x^{\prime \mu} = \{x^\mu, H_T \} & = &\left( \tau. x^\prime\right) \left( \tau^\mu+\frac{1}{m}h^{\mu\sigma} p_\sigma \right)\label{x}\\
  p'_\mu =\{p_\mu, H_T \} & = & \left( \tau. x^\prime\right) \left(-\partial_\mu \tau^\alpha p_\alpha -\frac{1}{2m}\partial_\mu h^{\rho\sigma}p_\rho p_\sigma \right)\label{p}
 \end{eqnarray}
Taking derivative of (\ref{x}) with respect to $\lambda$ and using (\ref{x}) and (\ref{p}) we get back the geodesic equation (\ref{geo}). Both Lagrangian and Hamiltonian analysis show that the geodesic equation is obeyed by the freely falling particle.

\subsection {\it{Gauge fixed analysis}}  
  
 So far we were  working in the gauge independent formalism. But to identify the physical variables
 and proceed with canonical quantization, gauge fixing is necessary.
 
 A particularly suitable choice of gauge is to identify the parameter $\lambda$ with the universal time,
 \begin{equation}
 \Omega_2 = x^0 - \lambda = 0
 \end{equation}
 As we shall see the Dirac brackets in this gauge are very simple and it is easy to identify the proper canonical variables of the theory.
 
 The constraints $\Omega_i$ and $\Omega_2$ now form a second class pair. The relevant matrix formed by the Poisson brackets of the two constraints is given by,
 \begin{equation}
 C_{ij} = \{\Omega_i, \Omega_j\} = -\Theta^{-1}\epsilon_{ij}, \,\,\, \epsilon_{12}=1
 \label{cmatrix}
 \end{equation}
 and its inverse is given by,
 \begin{equation}
 C_{ij}^{-1} = \Theta\epsilon_{ij}
 \label{cmatrixinverse}
 \end{equation}
 
 The Dirac brackets (denoted by a star)  between any two variables  are defined by,
 \begin{equation}
 \{f, g\}^* = \{f, g\} - \{f, \Omega_i \}C_{ij}^{-1} \{\Omega_j, g\}
 \label{diracbracket}
 \end{equation}
 Then the only non-vanishing Dirac brackets are given by,
 \begin{equation}
 \{x^\mu, p_\nu\}^* = \delta^\mu_\nu 
    - \delta_\nu^0 \Big(\tau^\mu +   \frac{1}{m}h^{\mu\sigma}p_\sigma \Big)\label{fdb}
 \end{equation}

 It is now possible to identify the physical variables of the system. There are eight phase space variables, two of which are eliminated by the constraints. That leaves us with six physical (phase space) degrees of freedom. We identify these with the set  $(x^i, p_j)$. Moreover their Dirac brackets are identical to the Poisson brackets,
 \begin{equation}
 \{x^i, p_j\}^* = \delta^i{}_j
 \label{canonical}
\end{equation}  
so that these may be regarded as a canonical pair. The variable $x^0$ is just the time parameter and, expectedly, has vanishing brackets with all variables. The other variable $p_0$ is eliminated in favour of the canonical set by using (\ref{pc1}). Realising that this constraint is now strongly implemented, we can solve for $p_0$ to get,

 \begin{eqnarray} 
 p_0 = -\Theta \left(\tau^i p_i + \frac{1}{2m} h^{ij}p_ip_j\right)
 \label{ham}
 \end{eqnarray}
 
 Finally, we have to identify the hamiltonian because the earlier expression (\ref{pc13}), based solely on the constraint, is now strongly zero. The new hamiltonian is given by,
 \begin{equation}
 H = - p_0
 \label{hamiltonian}
 \end{equation}
 
 To prove this fact we reproduce the equations of motion by taking the relevant Dirac brackets with the canonical variables. For example,

\begin{eqnarray}
{\dot{x}^i} = \{ x^i ,H \}^* = \Theta\left(\tau^i+ \frac{1}{m} h^{ij}p_j \right)   
\end{eqnarray} 
where (\ref{canonical}) is used. This matches exactly with $x^{\prime i}$ (\ref{x}) which has been obtained by gauge independent analysis, as $\tau.x^\prime  = \Theta$, since now we can put $\lambda = x^0 = t$, which is the gauge fixing constraint implemented strongly. Similarly, we can prove $\dot{p^i}$ matches with $p^{\prime i}$ in the gauge independent analysis. So $H$ generates the equations of motion of the physical variables.

\subsubsection{Canonical Quantization and  Schroedinger Equation}
Canonical quantization is now possible. We elevate the Dirac algebra (\ref{canonical}) to commutators, replacing the $x^i, p_j$ by operators. Then,
\begin{equation}
[\hat x^i, \hat p_j] = i \delta^i{}_j
\end{equation}
 
In the coordinate representation, therefore,
\begin{equation}
\hat x^i = x^i, \,\,\, \hat p_i  = -i\frac{\partial}{\partial x^i}
\label{rep}
\end{equation}
Using (\ref{rep}) along with (\ref{ham}) and (\ref{hamiltonian}), we can write down the following equation,
\begin{eqnarray}
i\frac{\partial \psi}{\partial t} = \Theta \left(\tau^i( -i\frac{\partial}{\partial x^i}) + \frac{1}{2m} h^{ij}(-\partial_i\partial_j)\right)\psi
\end{eqnarray}
This is the Schroedinger equation for a nonrelativistic particle in a Newton Cartan background.
As a consistency check we study its flat limit. In this limit $\Theta = 1,\quad \tau^i = 0,\quad h^{ij}=\delta^{ij}$ and the above equation reduces to,
\begin{equation}
i\frac{\partial \psi}{\partial t} = -\frac{1}{2m}\nabla^2 \psi
\end{equation}
Similarly, from (\ref{hamiltonian}) we find that  $H$ goes to its flat limit $\frac{p^2}{2m}$. These agreements are really wonderful.

\section{Conclusions}

The coupling of nonrelativistic matter to gravity is quite nontrivial when compared to the relativistic case. A prime reason is the lack of a single nondegenerate metric. While such a metric occurs naturally for relativistic theories, the nonrelativistic theory is saddled with a pair of degenerate metrics  \cite{Havas, Daut, TrautA, Kuch, EHL}. Thus it becomes necessary to adopt different techniques than those used conventionally. 

Among the various approaches, a large part is devoted  to gauging the Galilean (or its centrally extended ) algebra \cite{ABPR}. While this does reproduce the elements of the Newton Cartan geometry, it cannot illuminate the dynamics. Thus one takes recourse to introduce a background U(1) gauge field  to keep track with the symmetries in the dynamicsal
level.


For the specific problem of writing an action for a nonrelativistic free particle coupled to gravity, two methods were used in the literature. By the gauging of algebra method 
an action was written \cite{ PPP, j1}.
However, the particle equation of motion was not a geodesic \cite{PPP}. This last point brings us to the other approach \cite{Kuch}. Here the geodesic equation was {\it assumed}, from which an action was guessed.


Recently we have given a new form for the action of a nonrelativistic particle \cite{BM8} which was based on a systematic algorithm developed by us over the last few years to couple nonrelativistic theories to gravity \cite{BMM1,BMM2,BM4}. The algorithm is based on the gauge principle which dictates the precise method by which a  given  theory with global invariance can be reformulated into a theory with local invariance. New fields have to be introduced that help us to define covariant derivatives from the ordinary ones. The transformations of these new fields is fixed by requiring that the covariant derivatives transform in the same way under local transformations as the ordinary derivatives do under the global symmetry. Finally replacing the ordinary derivatives by the covariant derivatives in the original (flat space) action yields our cherished action which has an appropriate geometrical interpretation.  This action had a proper flat limit. No assumptions were used and the coupling to Newton Cartan geometry was simply an outcome of the method.

In this paper we have made a detailed canonical analysis of that action. Both lagrangian and hamiltonian formulations were discussed. We have shown that the Euler-Lagrange equation of motion is just the geodesic equation in the Newton Cartan background. This is a nontrivial check on the validity of our action, particularly since the issue of geodesy has been a recurring theme, both for the spinless \cite{PPP} and spinning particle models \cite{Berduci}.  We also like to mention that, contrary to \cite{Kuch},  we have not {\it assumed} the geodesic eqution. Rather, we have derived it from our action. 

The hamiltonian analysis of our model is based on Dirac's theory of constraints \cite{D}. This has been done using both gauge independent and gauge fixed versions. We find the appearance of a single first class constraint. The canonical hamiltonian vanishes, revealing  the reparametrisation invariance possessed by the action. This invariance is shown to be generated by the first class constraint. The time evolution of the system is given by the total hamiltonian, which is proportional to the first class constraint. Fixing this arbitrariness appropriately, we are able to reproduce the Euler Lagrange equations of motion. Thus the geodesic equation is also obtained in the hamiltonian formulation.

In the gauge fixed approach, we choose a gauge where the reparametrisation  parameter is taken to be the absolute time. The reparametrisation freedom is thus removed and the original first class constraint gets converted to second class. The Dirac brackets were computed. From these brackets the physical (canonical) variables were abstracted. The new hamiltonian was identified from which the original equations of motion were reproduced using the Dirac brackets. Thus the consistency of the gauge independent and gauge fixed formulations was established. Since the canonical pair had been abstracted it was possible to quantise the theory, say in the Schroedinger representation. The Schroedinger equation was explicitly written. All these results reproduced the expected flat limit. For instance, the Schroedinger equation found here smoothly goes over to the normal Schroedinger equation for a free nonrelativistic particle in flat spacetime.

   We have compared our result for the action with that found by other
approaches \cite{PPP}. The two expressions are apparently different but we
have explained the reasons for this feature. The origin of this
difference is contained in the respective flat space limits. Whereas we
consider a free NR particle in flat space and find its curved space
generalisation, the earlier approaches provide a covariant formulation
of an otherwise free NR particle in flat space but which is in the
presence of a Newtonian gravitational potential. Thus the final
covariant expressions in the two cases also differ. As we have further
shown, this difference cannot be `gauged'away by exploiting the Milne
symmetry.

As future possibilities, an immediate application would be to extend the analysis for a spinning particle. 
Such a model can be the starting point for more involved theories like the superparticle or even superstrings in a Newton Cartan background.

  .

\end{document}